\documentclass[twocolumn,prl,aps,superbib,tightenlines,floatfix]{revtex4}
\usepackage{amsfonts}
\usepackage{amsmath}
\usepackage{amssymb}
\begin{document}
\input epsf.sty
\topmargin -24mm
\textheight 243mm
\bibliographystyle{apsrev}
%\draft
%\wideabs
\title{Comment on ``First-principles treatments of electron transport properties for nanoscale junctions''}
\author{N.D. Lang$^a$ and M. Di Ventra$^b$}
\address{$^a$IBM Research Division, Thomas J. Watson Research Center, 
Yorktown Heights, NY 10598}
\address{$^b$Department of Physics, Virginia Polytechnic Institute and State University, 
Blacksburg, Virginia, 24061}
\maketitle
%]\bigskip
%\narrowtext

The recent paper by Fujimoto and Hirose~\cite{hirose} makes an unfortunate error in
discussing the use of the jellium model for the electrodes, which has the effect
of making it appear that this model is not adequate to treat the problem of
the conductance of gold nanowires. In fact it is entirely adequate, and
gives results quite similar to those found in the authors' more elaborate
treatment.

The main point is that the quantity $D$ discussed in Sec. III of Ref.~\onlinecite{hirose}, which represents
the distance between the jellium surface (positive-background edge) and the
plane of gold atoms (called the ``square basis of the nanowires'' in Ref.~\onlinecite{hirose}) 
contacting the nanowire is {\it not}, as Fujimoto and Hirose suggest, arbitrary or unknown,
but by construction of the jellium model, has a {\it perfectly definite value}. In
their case, this value is (1/4)a$_0$ = 1.93~a.u. (a$_0$ is the gold lattice
constant). If the correct value of $D=1.93$~a.u. is used in Fig. 6 of Ref.~\onlinecite{hirose}, which gives
conductance vs. $D$, then a conductance value of $\approx$ 0.98~G$_0$ is found, which
is to be compared with what the authors call the ``true'' value of $\approx$ 1~G$_0$.
This shows also that the effect on transport of the interface between the
jellium and the gold layer that seems to concern the authors is unimportant.

\begin{figure}
\epsfxsize=5cm \epsfbox{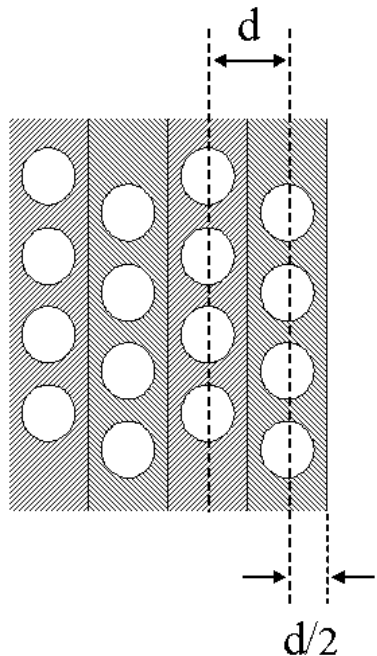}
\caption{Schematic of the construction of the jellium model out of an ionic lattice. The electrode is built 
up symmetrically slab by slab so that the positive-background edge of the jellium is 
half an interplanar spacing $d$ in front of the outermost lattice plane. Circles represent the ions and the 
jellium edge is on the right.}
\label{scheme}
\end{figure}

The essence of the construction of the jellium model is that {\it each} lattice
plane of the ions of the metal electrode is smeared out {\it symmetrically} 
into a uniform slab of positive charge (see Fig.~\ref{scheme}). Thus the plane of ions is at the
center of the slab which replaces it,
and the edge of the positive background is half an interplanar
spacing {\it in front} of the outermost lattice plane (see Fig.~\ref{scheme}).

In thinking about this construction, it is useful to consider the difference
$\delta V ({\bf r})$ between the potential due to the ionic lattice and the
semi-infinite positive background which represents it. This $\delta V ({\bf r})$, which can
be used to perturbatively re-introduce the discrete lattice as done e.g. in
Ref.~\onlinecite{lang1}, can only be viewed as a small perturbation if the sheet of ions is at
the slab center, since otherwise $\delta V ({\bf r})$ will not vanish at 
$\pm \infty$ (see footnote 25 of Ref.~\onlinecite{lang2}). These issues are also discussed in
Ref.~\onlinecite{lang3}.

Since for Au(100), discussed in Ref.~\onlinecite{hirose}, the interplanar spacing is
(1/2)a$_0$, the distance between the jellium surface and a gold layer put down
on it will be half of this, i.e. (1/4)a$_0$. This spacing, as noted before,
gives the correct conductance. Similar conclusions can be drawn for other systems as well. 
For instance, it was shown in Ref.~\onlinecite{diventra} that the conductance of an organic molecule between 
two jellium electrodes is quite similar whether the molecule makes contact directly with the jellium 
surface or with a plane of gold atoms adjacent to the jellium surface. 

One of us (MD) acknowledges support from the NSF Grant Nos. DMR-01-02277 and
DMR-01-33075, and Carilion Biomedical Institute.

\end{document}